\documentclass[final,5p,times,twocolumn]{article}

\usepackage[utf8]{inputenc}
\usepackage{natbib}
\usepackage{graphicx}
\usepackage[margin=0.5in]{geometry}

\usepackage{lineno,hyperref}
\modulolinenumbers[5]

\usepackage{epstopdf}
\usepackage{verbatim}
\usepackage{mathrsfs}
\usepackage{array}

\newcolumntype{H}{>{\setbox0=\hbox\bgroup}c<{\egroup}@{}}

\begin{document}

\title{Large-scale asymmetry in the distribution of galaxy spin directions -- analysis and reproduction}
\date{}

\author{Lior Shamir\footnote{lshamir@mtu.edu}  \\ Kansas State University \\ 1701 Platt St \\ Manhattan, KS 66506, USA}

\maketitle

\abstract{
Recent independent observations using several different telescope systems an analysis methods have provided evidence of parity violation between the number of galaxies that spin in opposite directions. On the other hand, other studies argued that no parity violation can be identified. This paper provides detailed analysis, statistical inference, and reproduction of previous reports that show no preferred spin direction. Code and data used for the reproduction are publicly available. The results show that the data used in all of these studies agrees with the observation of a preferred direction as observed from Earth. In some of these studies the datasets were too small, or the statistical analysis was incomplete. In other papers the results were impacted by experimental design decisions that lead directly to show non-preferred direction. In some of these cases these decisions are not stated in the papers, but were revealed after further investigation in cases where the reproduction of the work did not match the results reported in the papers. These results show that the data used in all of these previous studies in fact agree with the contention that galaxies as observed from Earth have a preferred spin direction, and the distribution of galaxy spin directions as observed from Earth form a cosmological-scale dipole axis. This study also shows that the reason for the observations is not necessarily an anomaly in the large-scale structure, and can also be related to internal structure of galaxies.}

\section{Introduction}
\label{introduction}

According to the Cosmological Principle, the Universe is expected to be homogeneous and isotropic. On the other hand, a large number of studies using numerous different probes show cosmological-scale isotropy and parity violations, suggesting that the Cosmological Principle might not be aligned with observations \citep{aluri2022observable}. In addition to the cosmic microwave background radiation \citep{lue1999cosmological,eriksen2004asymmetries,abramo2006anomalies,cline2003does,gordon2004low,land2005examination,campanelli2007cosmic,mariano2013cmb,ade2014planck,zhe2015quadrupole,santos2015influence,dong2015inflation,gruppuso2018evens,ashtekar2021cosmic,luongo2022larger,yeung2022directional}, such probes also include dark energy \citep{adhav2011kantowski,adhav2011lrs,perivolaropoulos2014large,colin2019evidence}, $H_o$ anisotropy \citep{luongo2021larger,dainotti2022evolution}, distribution of galaxy morphology types \citep{javanmardi2017anisotropy}, cosmic rays \citep{aab2017observation}, quasars \citep{hutsemekers2014alignment,slagter2022new,quasars,zhao2021tomographic,secrest2021test}, LX-T scaling \citep{migkas2020probing}, Ia supernova \citep{javanmardi2015probing,lin2016significance}, short gamma ray bursts \citep{meszaros2019oppositeness}, large-scale clustering of {\it Fermi} blazars \citep{marcha2021large}, radio galaxies \citep{taylor2016alignments,contigiani2017radio,panwar2020alignment}, and the tetrahedral arrangements of sets of four galaxies \citep{philcox2022probing,hou2022measurement}.

Large-scale cosmological structures that shift from the Cosmological Principle and the Friedmann–Lemaître–Robertson–Walker (FLRW) Universe \citep{aluri2022observable} can be related to ``alternative'' theories that were proposed in the past few decades \citep{aluri2022observable}. These theories include contraction prior to inflation \citep{piao2004suppressing}, spinor-driven inflation \citep{bohmer2008cmb}, moving dark energy \citep{jimenez2007cosmology}, primordial anisotropic vacuum pressure \citep{rodrigues2008anisotropic},  spin foam cosmology \citep{rovelli2010spinfoam,kisielowski2012one}, multiple vacua \citep{piao2005possible}, inhomogeneous Big Bang singularity \citep{schneider1999models}, cosmological big bounce \citep{battisti2010big}, double inflation \citep{feng2003double}, ellipsoidal universe \citep{campanelli2006ellipsoidal,campanelli2007cosmic,campanelli2011cosmic,gruppuso2007complete,cea2014ellipsoidal}, longitudinal gravitational wave cosmology \citep{mol2011gravitodynamics}, anisotropic dark energy \citep{adhav2011kantowski,adhav2011lrs}, or rotating universe \citep{godel1949example,ozsvath1962finite,ozsvath2001approaches,sivaram2012primordial,chechin2016rotation,seshavatharam2020integrated,camp2021}.  Dipole cosmological models that are based on the $\Lambda CDM$ cosmology expanded to support a dipole universe \citep{krishnan2023dipole} and a dipole big bang \citep{allahyari2023big} were also proposed.

The contention of charge conjugation parity (CP) symmetry violation had been proposed in particle physics as early as the 1950's \citep{lee1956question}, and clear evidence of CP violation was provided in the 1960s \citep{cronin1964evidence}. These surprising observations can have implications on parity violation and dipoles in the Universe \citep{hou2011source,bian2018gravitational,gava2010cp,flanz1995baryogenesis}. Possible charge conjugation, parity transformation, and time reversal (CPT) symmetry violation \citep{lehnert2016cpt} has also been proposed as a factor linked to an anisotropic or polarized Universe \citep{crowder2013measurement,mavromatos2017models,mavromatos2018spontaneous,poplawski2011matter}.

One of the cosmological models aligned with the contention that the Universe is oriented around a large-scale axis is black hole cosmology \citep{pathria1972universe}, according which the Universe is the interior of a black hole in another universe \citep{pathria1972universe,stuckey1994observable,easson2001universe,seshavatharam2010physics,poplawski2010radial,tatum2018flat,christillin2014machian,chakrabarty2020toy}. Since black holes spin \citep{gammie2004black,takahashi2004shapes,volonteri2005distribution,mcclintock2006spin,mudambi2020estimation,reynolds2021observational}, a universe hosted inside a black hole should inherit its spin from its host black hole \citep{poplawski2010cosmology,seshavatharam2010physics,seshavatharam2014understanding,christillin2014machian,seshavatharam2020light,seshavatharam2020integrated}.  

If our Universe is the interior of a black hole in another universe, it agrees with the theory of multiverse \citep{carr2008universe,hall2008evidence,antonov2015hidden,garriga2016black,debnath2022anisotropic,trimble2009multiverses,kragh2009contemporary}. Black hole cosmology is also supported by the agreement between the Schwarzschild radius of the Universe and the Hubble radius \citep{christillin2014machian}, as well as the accelerated expansion of the Universe without the need to assume the existence of dark energy. A black hole universe is also expected to violate the CPT symmetry \citep{lehnert2016cpt}. Black hole cosmology is directly related to the ability to view space as a projection, which is the theory of holographic universe \citep{susskind1995world,bak2000holographic,bousso2002holographic,myung2005holographic,hu2006interacting,rinaldi2022matrix,sivaram2013holography,shor2021representation}.

Quantum Field Theory (QFT) has shown promising success in explaining microscopic phenomena such as sub-atomic structures \citep{davies1976quantum}. Early analysis predicted an anisotropic universe, noting that ``It is a curious unexplained feature that the present
condition of the Universe is one of high isotropy'' \citep{davies1976quantum}. Given QFT, cosmological-scale gravitational dipoles can be expected \citep{faulkner2022snowmass}. Such gravitational dipoles are driven by matter and antimatter that have gravitational charges, and can explain phenomena normally attributed to dark energy and dark matter, and does not assume primordial singularity or cosmic inflation \citep{hajdukovic2013signatures,hajdukovic2014virtual}. The quantum vacuum can then be explained by polarized gravity, and the gravitational dipoles can also explain disagreement between the observed and theoretically predicted dark energy density \citep{hajdukovic2013signatures,hajdukovic2014virtual}. The poles as gravitational sources in the Universe can change the large-scale symmetry, provoking fluctuations that contribute in the evolution of the Universe and its geometrical aspect, making its asymmetricity possible.

As {\it tidal torque theory} can explain the initial origin of galaxy angular momentum \citep{hoyle1949problems,peebles1969origin,doroshkevich1970spatial,porciani2002testing2,schafer2009galactic}, it has also been associated with the large-scale structure, and a link between the alignment of galaxy rotation and the large-scale structure has been proposed \citep{schafer2009galactic}. Empirical evidence of a link between the large-scale structure and the alignment of galaxy spin directions was reported using the Sloan Digital Sky Survey (SDSS) and the Two-Micron All-Sky Survey \citep{jones2010fossil}, and such link was reported by other consequent studies \citep{jones2010fossil,tempel2013evidence,tempel2013galaxy,tempel2014detecting,pahwa2016alignment,ganeshaiah2018cosmic,lee2018wobbling,ganeshaiah2019cosmic,lee2019galaxy,lee2019mysterious,blue2020chiles,welker2020sami,kraljic2021sdss,lopez2021deviations,motloch2021observed}.

In addition to the probes mentioned above, another probe that supports the contention of anisotropic Universe and a preferred direction is the asymmetric distribution of galaxies with opposite spin directions. Early reports of the asymmetry used a small number of manually annotated galaxies in the local supercluster, suggesting that the number of galaxies spinning clockwise is different from the number of galaxies spinning counterclockwise with certainty of 92\% \citep{macgillivray1985anisotropy}. 

The deployment of robotic telescopes allowed the analysis of a larger number of galaxies. Perhaps the first major modern high-throughput autonomous digital sky survey was SDSS \citep{york2000sloan}, with data acquisition power that was unprecedented at the time. Analysis of the distribution of the spin directions of spiral galaxies in SDSS images showed parity violation and a dipole axis, observed in several different experiments using manual and automatic annotation \citep{longo2011detection,shamir2012handedness,shamir2020patterns,shamir2020pasa,shamir2021particles,shamir2021large,shamir2022large,shamir2022analysis2}. In addition to SDSS, experiments with other sky surveys also showed similar large-scale parity violation in galaxy spin directions as observed from Earth. These experiments include data from Hubble Space Telescope \citep{shamir2020pasa}, Pan-STARRS \citep{shamir2020patterns}, DECam \citep{shamir2021large}, DES \citep{shamir2022asymmetry}, and DESI Legacy Survey \citep{shamir2022analysis2}. Some of the experiments included over $10^6$ galaxies \citep{shamir2022analysis2}, providing strong statistical significance. The experiments showed clear separation between hemispheres, such that one hemisphere has an excessive number of galaxies spinning clockwise, while the opposite hemisphere contained more galaxies that seem to spin counterclockwise to an Earth-based observer. An analysis with $\sim1.3 \cdot 10^6$ galaxies allowed to show a dipole axis alignment without the need to fit it in a certain statistical model \citep{shamir2022analysis2}. The parity violation seems to become stronger as the redshift increases \citep{shamir2020patterns,shamir2022large}. Other studies include a smaller number of galaxies to show links between the spin directions of neighboring galaxies \citep{mai2022sami}, including galaxies that are too far {\bf from each other} to have gravitational interactions \citep{lee2019mysterious}. These links were defined ``mysterious'', suggesting that galaxies in the Universe are connected through their spin directions \citep{lee2019mysterious}. A correlation was also identified between the cosmic initial conditions and spin directions of galaxies \citep{motloch2021observed}. Experiments showed that the strength of the asymmetry is not necessarily affected when limiting the size of the galaxies \citep{shamir2020patterns}, but showed inconclusive evidence of 1.1$\sigma$ that the asymmetry increases as the density of the galaxies gets larger \citep{shamir2022large}.

While these studies showed patterns of alignment in galaxies spin directions at scales far larger than any known astrophysical structure, numerous other studies showed alignment in the spin directions of galaxies at smaller scales. For instance, a set of 1,418 galaxies from the Sydney-Australian-Astronomical-Observatory Multi-object Integral-Field Spectrograph (SAMI) survey \citep{bryant2015sami,croom2021sami} showed spin alignment of galaxies within filaments \citep{welker2020sami}, supported by a consequent study that also showed a link between spin alignment and the morphology of the galaxies \citep{barsanti2022sami}. Spin direction alignment in cosmic web filaments has also been shown by multiple other studies \citep{tempel2013evidence,tempel2013galaxy,kraljic2021sdss}, as well as in numerical simulations \citep{zhang2009spin,davis2009angular,libeskind2013velocity,libeskind2014universal,forero2014cosmic,wang2018build,lopez2019deviations}.

But while several studies showed non-random distribution of galaxy orientation, some experiments argued that the distribution is random \citep{iye1991catalog,land2008galaxy,hayes2017nature,iye2020spin}.  One of the first documented experiments to claim for random distribution and no preferred handedness in the distribution of galaxy spin directions was an experiment based on manually collected galaxies that took place in the pre-information era in astronomy \citep{iye1991catalog}. During that time, large-scale autonomous digital sky surveys did not yet exist, which limited the ability to collect and analyze large databases within reasonable efforts. The dataset included 3,257 galaxies annotated as spinning clockwise, and 3,268 galaxies annotated as spinning counterclockwise. The downside of the study was that due to the limited ability to collect data at the time, the dataset was relatively small. The small size of the dataset does not allow to provide statistical significance given the magnitude of the asymmetry reported here and in previous reports.

For instance, the asymmetry between the number of clockwise and counterclockwise galaxies as described in \citep{shamir2020patterns} is $\sim$1.4\%. Given that asymmetry, showing a one-tailed P value of 0.05 requires 55,818 galaxies. Therefore, the dataset of a few thousand galaxies is not large enough to show a statistically significant asymmetry. 
But while these experiments use a small number of galaxies, other experiments used larger datasets, but still showed random distribution of the galaxy spin directions.

The purpose of this paper is to carefully examine the claims and experiments, and understand the conflicting results between different experiments that in some cases use the same data but reach different conclusions.  Section~\ref{previous_studies} discusses the results provided by the {\it Galaxy Zoo} citizen science initiative to annotate galaxies by their spin directions \citep{land2008galaxy}, Section~\ref{sparcfire_analysis} reproduces the analysis of SDSS galaxies with spectra annotated by the {\it SpArcFiRe} method \citep{hayes2017nature}, Section~\ref{reproduction} reproduces an experiment of SDSS galaxies annotated using the {\it Ganalyzer} method \citep{iye2020spin}, Section~\ref{error} discusses possible reasons for the observations, and Section~\ref{explanation} provides a possible explanation to the observation that does not necessarily require to shift from the standard model.

\section{Early analyses with manual annotation and Galaxy Zoo crowdsourcing}
\label{previous_studies}



Another notable experiment was based on galaxies annotated manually by a large number of volunteers through the {\it Galaxy Zoo} platform \citep{land2008galaxy}. The analysis by using manual annotation was limited by a very high error rate, but after selecting just the galaxies on which the rate of agreement was high, the error can be reduced. But more importantly, the annotations were systematically biased by the human perception or the user interface \citep{land2008galaxy}. Even when using the ``superclean'' criterion, according which only galaxies on which 95\% or more of annotations agreed are used, the asymmetry was $\sim$15\%. That asymmetry was determined to be driven by the bias rather than a reflection of the real distribution of galaxies in the sky \citep{land2008galaxy}.

When the bias was noticed, a new experiment was done by annotating a small subset of 91,303 galaxies using the same platform. In addition to annotating the original images, the second experiment also included the annotation of the mirrored image of each galaxy. The annotation of the mirrored images is expected to offset the bias. 

After selecting the ``sueprclean'' annotations, the results showed 5.525\% clockwise galaxies compared to 5.646\% mirrored counterclockwise galaxies that were annotated as clockwise. The information is specified in Table 2 in \citep{land2008galaxy}. That showed a 2.1\% higher number of counterclockwise galaxies. Similar results were also shown with galaxies annotated as spinning counterclockwise, where 6.032\% of the non-mirrored images were annotated as spinning counterclockwise, while 5.942\% of the mirrored galaxies were annotated as clockwise. The magnitude and direction of the 1\%-2\% asymmetry is aligned with the asymmetry reported in \citep{shamir2020patterns}, which is also based on SDSS galaxies with spectra, and therefore the footprint and limiting magnitude of the galaxies used in \citep{land2008galaxy} and in \citep{shamir2020patterns} are expected to be similar. 

The primary difference between the experiment of \citep{land2008galaxy} and the experiment of \citep{shamir2020patterns} is the size of the datasets. While in \citep{shamir2020patterns} more than $6\cdot10^4$ galaxies were used, the set of ``superclean'' Galaxy Zoo galaxies that were annotated as mirrored and non-mirrored to offset for the human bias was just $\sim10^4$. Table~\ref{galaxy_zoo_table} shows the number of galaxies that spin clockwise and counterclockwise as annotated by Galaxy Zoo when the original images were annotated, and when the mirrored images were annotated. 

\begin{table}
\centering
\scriptsize
\begin{tabular}{|l|c|c|c|c|}
\hline
                   & Original          & Mirrored   & $\frac{\#ccw}{\#cw}$        & P                      \\
                   &                      &                &                                    & (one-tailed)       \\
\hline
Clockwise             & 5,044      & 5,155       &   1.022                          &    0.13   \\
Counterclockwise  &  5,507     & 5,425      &   1.015                          &    0.21    \\ 
\hline
\end{tabular}
\caption{The number of galaxies in the original images and mirrored image annotated as spinning clockwise or counterclockwise through Galaxy Zoo \citep{land2008galaxy}.}
\label{galaxy_zoo_table}
\end{table}

While the crowdsourcing analysis provided more than $10^5$ annotated galaxies, it is still not of sufficient size to show statistically significant asymmetry. On the other hand, it does show a consistent higher number of counterclockwise galaxies compared to clockwise galaxies. For instance, the number of galaxies annotated as clockwise is lower than the number of mirrored galaxies annotated as clockwise, which is 5,044 and 5,155, respectively. That provides a P$\simeq$0.13 one-tailed statistical significance of the binomial distribution assuming the probability of each spin direction is 0.5. 

The numbers of galaxies annotated as counterclockwise and the number of mirrored galaxies annotated as counterclockwise were 5,507 and 5,425, respectively, providing a binomial distribution statistical signal of $P\simeq0.21$. That asymmetry cannot be considered statistically significant, but it also agrees in direction and magnitude with the asymmetry shown in  \citep{shamir2020patterns} for the same footprint and distribution in the sky. It is possible that if the dataset used in \citep{land2008galaxy} was larger the results would have been statistically significant, but since the dataset is of a limited size that assumption cannot be proven or disproven. 

While none of the two experiments show statistical significance, they both show a higher number of galaxies that spin counterclockwise in the SDSS footprint. The aggregated P-value of the two experiments is 0.0273. Although the annotations of each galaxy are different, some of the galaxies might exist in both experiments, where in one experiment the galaxy was annotated using its original image, and in the other experiment it was annotated using its mirrored image. The presence of galaxies that were used in both experiments therefore does not allow to soundly aggregate the P values. But in any case, even if the P value does not show statistical significance, the results observed with Galaxy Zoo data certainly do not conflict with the results shown in \citep{shamir2020patterns} for SDSS galaxies with spectra.

\section{A previous experiment using Galaxy Zoo galaxies annotated by {\it SpArcFiRe}}
\label{sparcfire_analysis}

Another analysis \citep{hayes2017nature} of SDSS galaxies with spectra used an automatic annotation to provide an analysis with a far larger number of galaxies compared to \citep{land2008galaxy}. The SDSS galaxies were the galaxies also used by {\it Galaxy Zoo 1}. To separate the galaxies by their spin direction, the {\it SpArcFiRe} method \citep{hayes2014} was used. {\it SpArcFiRe} (SPiral ARC FInder and REporter) receives a galaxy image as an input, normally in the PNG image file format, and extract several descriptors of the galaxy arms. The algorithm works by first identifying arm segments in the galaxy image, and then grouping the pixels that belong in each segment. Once the pixels the different arm segments are identified, the pixels of each arm segment are fitted to a logarithmic spiral arc. The fitness of the pixels in the arm segment provides information about the arm, that can also be used to identify its curve direction, and consequently the spin direction of the galaxy. Source code of the implementation of the {\it SpArcFiRe} algorithm is available at \url{https://github.com/waynebhayes/SpArcFiRe}, and a full detailed description of the method is available at \citep{hayes2014}. Processing of a 128$\times$128 galaxy galaxy image takes $\sim$30 seconds using an Intel Core-i7 processor, and therefore a set of 100 cores was used to annotate the $\sim6.7\cdot10^5$ galaxies in the {\it Galaxy Zoo 1} dataset. 

After applying the {\it SpArcFiRe}  method to separate the galaxies by their spin direction, the results of the study showed that when separating the spiral galaxies from the elliptical galaxies through Galaxy Zoo 1 annotations, the asymmetry can be considered statistically significant. As shown in Table 2 in \citep{hayes2017nature}, the statistical significance ranged between 2$\sigma$ to 3$\sigma$, depends on the agreement threshold of the {\it Galaxy Zoo} separation between the elliptical and spiral galaxies. These results agree with the contention of a higher number of galaxies spinning counterclockwise in that footprint, but might be biased due to a possible human bias in the selection of spiral galaxies. For instance, if the human annotators tend to select more counterclockwise galaxies as spiral, that can lead to an observed asymmetry when annotating the spin directions of these galaxies \citep{hayes2017nature}.

To avoid the bias in the human selection of spiral galaxies, another experiment was made by selecting the spiral galaxies by using a machine learning algorithm. Naturally, the machine learning algorithm was trained by spiral and elliptical galaxies. To avoid bias in the classification, the class of spiral galaxies included the same number of clockwise and counterclockwise galaxies \citep{hayes2017nature}. However, in addition to having a balanced number of galaxies in each class, the machine learning algorithm was limited to features that cannot identify the spin direction of the galaxy. That is, all features that showed a certain correlation with the spin direction of the galaxy were removed manually and were not used in the analysis. As stated in \citep{hayes2017nature}, ``We choose our attributes to include some photometric attributes that were disjoint with those that Shamir (2016) found to be correlated with chirality, in addition to several SpArcFiRe outputs with all chirality information removed''.

When manually removing the features that correlate with the asymmetry, it can be expected that the step of separation of the galaxies by their spin directions would provide a lower asymmetry signal. To test that empirically, the same experiment was reproduced with the same code, but without using machine learning to identify spiral galaxies, and therefore also without manually removing specific features. Two ways of selecting the galaxies were used. The first was by not applying any selection of spiral galaxies. That is, the SpArcFiRe algorithm was applied to all Galaxy Zoo 1 galaxies, and all galaxies that SpArcFiRe was able to determine their spin directions were used in the analysis \citep{mcadam2023reanalysis}. That led to a dataset of 271,063 galaxies. The other way for selecting spiral galaxies was by applying the Ganalyzer algorithm \citep{shamir2011ganalyzer} to select spiral galaxies, but without separating the galaxies by their spin direction.

Ganalyzer works by first converting the galaxy image into its radial intensity plot, which is a $360\times35$ image such that the value of the pixel at Cartesian coordinates $(x,y)$ is the value of the pixel in the polar coordinate $(\theta,r)$ in the original galaxy image, where the centre point is the center of the galaxy. The value of $\theta$ is within $(0,360)$, and the radius r is in percentage of the galaxy radius. Then, a peak detection algorithm is applied to identify the maximum brightness in each line of the radial intensity plot. Since galaxy arms are brighter than the background pixels at the same radial distance from the galaxy center, the peaks are the galaxy arms. If the peaks are aligned in straight vertical lines, it means that the angle of the arm does not change with the radial distance, and therefore the galaxy is not spiral. But if the peak form a line that is not vertical, it shows that the arms are spiral. The algorithm is explained in full detail and experimental results in \citep{shamir2011ganalyzer}. 

\begin{figure*}
\includegraphics[scale=0.27]{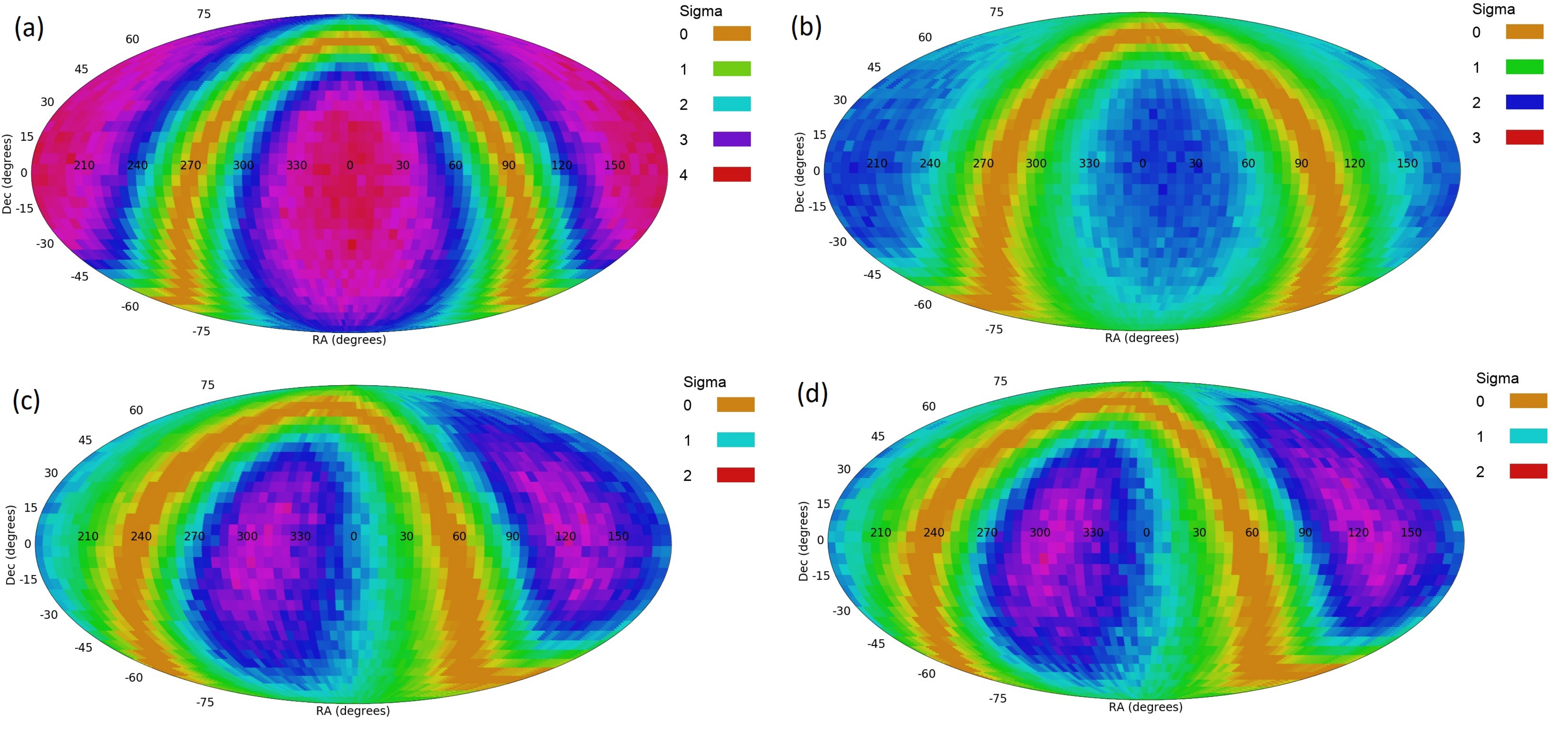}
\caption{Results of analysis of Galaxy Zoo 1 galaxies annotated by SpArcFiRe. Panel (a) is the result of the analysis with spiral galaxies selected by Ganalyzer when the images are mirrored, Panel (b) shows the analysis when the galaxies are not mirrored. Panels (c) and (d) show the results of the analysis without a first step of selection of spiral galaxies when the galaxy images are mirrored or non-mirrored, respectively.}
\label{sparcfire}
\end{figure*}

Ganalyzer is not based on machine learning or pattern recognition, and therefore does not have a step of training or feature selection. The dataset of annotated galaxies after selecting spiral galaxies contained 138,940 galaxies. Both datasets can be accessed at \url{https://people.cs.ksu.edu/~lshamir/data/sparcfire}. Figure~\ref{sparcfire} shows the results of the analysis when using the original galaxy images and the mirrored galaxy images.

As the figure shows, the results are different when using the original images and the mirrored images. That is expected due to the asymmetric nature of the SpArcFiRe software is discussed in the appendix of \citep{hayes2017nature}. The statistical significance of the datasets ranges from 2.05$\sigma$ when using the non-mirrored images without a first step of spiral selection, to 3.6$\sigma$ when using the mirrored images when using the mirrored galaxy images annotated after a first step of selecting spiral galaxies. 



\section{Previous experiment by using deep neural networks for annotating the galaxies}
\label{deep_learning}

In the past decade, deep neural networks became a very common tool for annotating galaxies by their morphology \citep{dieleman2015rotation,graham2019galaxy,mittal2019data,hosny2020classification,cecotti2020rotation,cheng2020optimizing}. Neural networks have been shown to provide superior performance in their ability to annotate images automatically, and due to the availability of libraries such as TensorFlow or PyTorch their implementation is normally more accessible compared to model-driven approaches. The downside of deep neural networks is that they work by complex data-driven rules, which makes them non-trivial to understand. That can add many unexpected biases that makes neural network imperfect for detecting subtle asymmetries reflected by the annotation of image data \citep{dhar2022systematic}.

An application of a deep neural network to annotate galaxies by their sin direction was performed in \cite{tadaki2020spin}, annotating HCS galaxies by using deep convolutional neural networks. The annotation provided 38,718 galaxies spinning clockwise, and 37,917 spinning counterclockwise in the HCS footprint. The higher number of clockwise galaxies in the HCS footprint is in agreement with analysis using the DESI Legacy Survey, also showing a higher number of galaxies spinning clockwise around that part of the sky \citep{shamir2022analysis2}. The one-tailed probability to such distribution to occur by chance is P=0.0019. Because the biases of deep neural networks are very difficult to control \citep{rudin2019stop,dhar2022systematic}, such analysis cannot provide a clear proof of non-random distribution of galaxy spin directions, as also state in the paper \citep{tadaki2020spin}. But the differences as reported in \citep{tadaki2020spin} also definitely do not conflict with the contention that the distribution of galaxy spin directions is not necessarily symmetric, and in fact agree with that contention rather than disagree with it.

\section{Reproduction of analysis of 72,888 SDSS galaxies}
\label{reproduction}


Another experiments that argued for no preferred handedness in the distribution of galaxy spin directions using a relatively large number of galaxies is \citep{iye2020spin}. The analysis of \cite{iye2020spin} is based on a dataset of 162,514 photometric objects used in \citep{shamir2017photometric}. The main argument of the work suggests that the dataset contains a large number of ``duplicate objects'', and once these objects are removed the dataset does not show non-random distribution \citep{iye2020spin}.

The dataset used in \citep{shamir2017photometric} was used for photometric analysis of objects that spin in opposite directions. The study \citep{shamir2017photometric} does not make any claim for the presence or absence of any kind of axis in that data, and no such claim about that dataset was made in any other paper. While several previous experiments were made with SDSS galaxies to show a dipole axis formed by the distribution of galaxy spin directions  \citep{shamir2012handedness,shamir2020patterns,shamir2021particles}, none of these experiments were based on the dataset used in \citep{shamir2017photometric}. When using the photometric objects used in \citep{shamir2017photometric} to study the distribution of galaxy spin directions, photometric objects that are part of the same galaxies indeed become ``duplicate objects''. But as mentioned above, \citep{shamir2017photometric} does not make any claim for the presence of any kind of axis, and no such claim was made about that dataset in any other paper.

Although the dataset used in \citep{shamir2017photometric} was not used in previous papers to analyze a dipole axis, it is still expected that the dataset used in \citep{shamir2017photometric} would be consistent with the results shown by previous datasets. The ``clean'' dataset used by \cite{iye2020spin} was compiled by removing duplicate objects from the dataset used in \citep{shamir2017photometric}. That provided a dataset of 72,888 galaxies \citep{iye2020spin}. That dataset is available for download at \url{https://people.cs.ksu.edu/~lshamir/data/iye_et_al/galaxies.csv}.

As explained in Equation 2 in Section 2.1 of \citep{iye2020spin}, the strength of the dipole D from a certain point in the sky is determined by $\Sigma_{i=1}^N h^i \Omega^i P / N = \Sigma_{i=1}^N h^i cos \theta^i / N $,
where $P$ is the fiduciary pole vector, $\Omega^i$ is the spin vector of galaxy $i$, $h^i$ is the spin direction of galaxy $i$, $\theta^i$ is the angle between the direction of the galaxy $i$ and the direction of the pole vector P, and $N$ is the total number of galaxies in the dataset.

The spin direction $h^i$ of the galaxy $i$ is within the set $\{1,-1\}$. The statistical strength of a dipole to exist at a certain point in the sky was determined by the D when the spin directions of the galaxies were taken from the dataset, compared to the mean $\overline{D}$ and standard deviation $D_{\sigma}$ computed after running the same analysis when assigning the galaxies with random spin directions. The full implementation of the method is available at \url{https://people.cs.ksu.edu/~lshamir/data/iye_et_al}.

To follow the experiment of \cite{iye2020spin}, the mean and standard deviation of D when using random spin directions was done 50,000 times, although the change in the results becomes minimal after 2,000 runs. Following the description of \citep{iye2020spin}, the spin direction of the galaxy d is taken from the same dataset of 72,888 galaxies used in \citep{iye2020spin}, and also available publicly at \url{https://people.cs.ksu.edu/~lshamir/data/iye_et_al/galaxies.csv}.



As also done in \citep{iye2020spin}, the location of the most likely dipole axis was determined by the statistical $\sigma$ difference between the D computed with the spin directions of the galaxies and the D computed when the galaxies are assigned with random spin directions determined the most likely position and the statistical signal of the dipole. The code that implements the method and step-by-step instructions to reproduce the results are available at \url{https://people.cs.ksu.edu/~lshamir/data/iye_et_al}.

\subsection{Results}
\label{results}

\cite{iye2020spin} performed experiments with two datasets. The first was the entire dataset used in \citep{shamir2017photometric}. The second dataset was a subset of the dataset used in \citep{shamir2017photometric}, such that the redshift of the galaxies was limited to less than 0.1. For instance, the low statistical significance of 0.29$\sigma$ reported in the abstract of \citep{iye2020spin} as the statistical significance of the sample was in fact the statistical significance observed in the sub-sample of galaxies with redshift lower than 0.1, as shown in Table 1 in \citep{iye2020spin}. 

The low statistical significance when limiting the redshift was reported previously in \citep{shamir2020patterns}. For instance, Tables, 3, 5, 6 and 7 in \citep{shamir2020patterns} show random distribution when the redshift is limited to 0.1. Also, an experiment reported in \citep{shamir2020patterns} used galaxies limited to $z<0.15$, and showed that the statistical significance of the dipole axis was below 2$\sigma$. A similar experiment \citep{shamir2022large} also showed that the dipole axis is not statistically significant in the lower redshift ranges, including $0<z<0.1$. Therefore, limiting the redshift to 0.1 is expected to show low statistical significance. The low statistical significance shown in \citep{iye2020spin} agrees with the random distribution in that redshift range as shown in \citep{shamir2020patterns,shamir2022large}.

When not limiting the redshift, \cite{iye2020spin} argue that the statistical significance of the ``clean'' dataset of 72,888 galaxies exhibits a dipole axis in the galaxy spin directions with statistical significance of 1.29$\sigma$. That is specified in Table 1 in \citep{iye2020spin}. However, the reproduction of the experiment using the exact same 72,888 galaxies and the code described in Section~\ref{reproduction} shows that the statistical significance of the dipole axis is 2.15$\sigma$. The code, data, step-by-step instructions, and the output of the analysis are provided at \url{https://people.cs.ksu.edu/~lshamir/data/iye_et_al}.

\begin{figure*}
\centering
\includegraphics[scale=0.45]{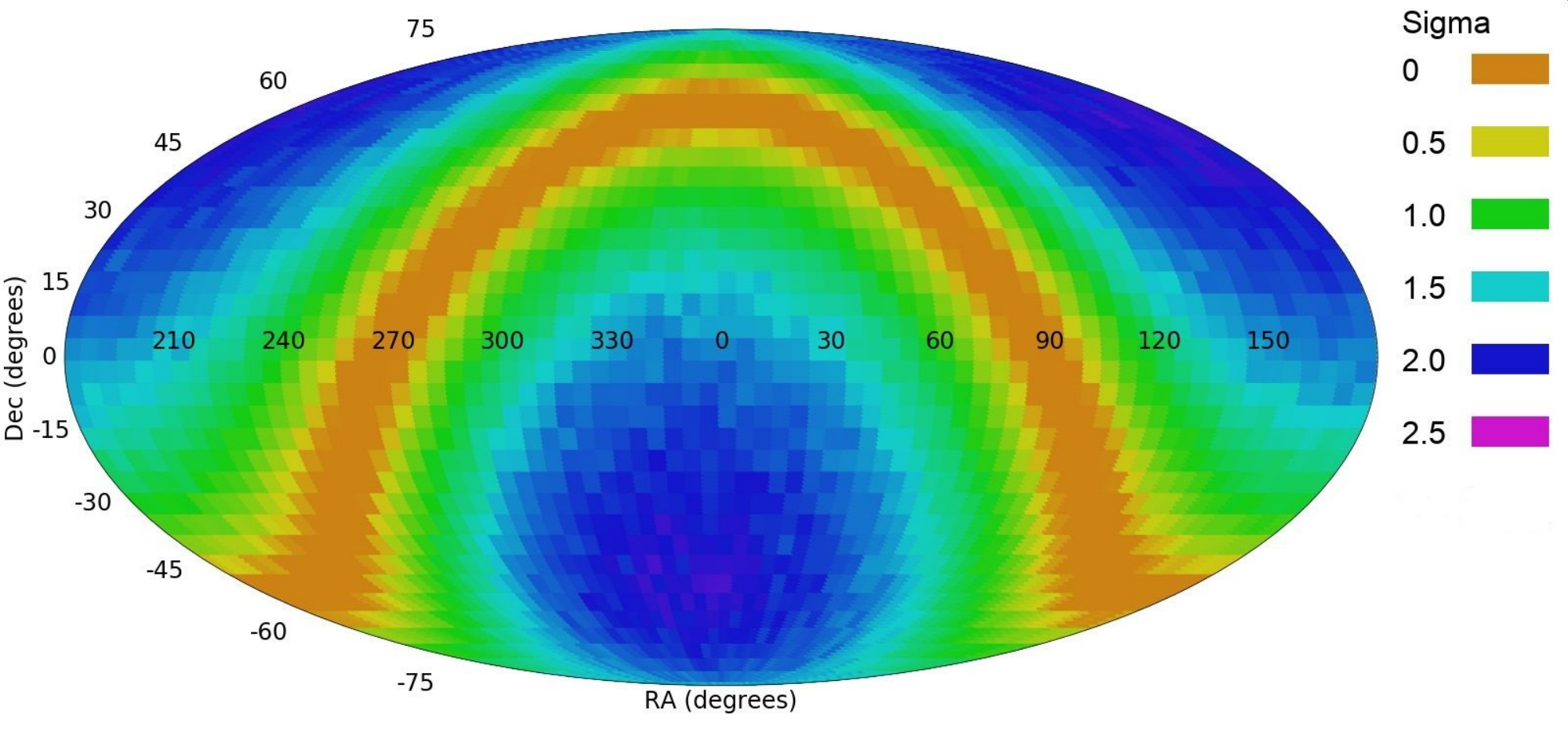}
\caption{The statistical significance for a dipole axis to exist by chance from all $(\alpha,\delta)$ combinations. Reproduction of the analysis is available at \url{https://people.cs.ksu.edu/~lshamir/data/iye_et_al}.}
\label{dipole1}
\end{figure*}

Figure~\ref{dipole1} provides a Mollweide projection that visualizes the statistical significance of a dipole axis from every possible integer combination of $(\alpha,\delta)$. The most likely dipole axis with statistical significance of 2.15$\sigma$ is observed at $(\alpha=170^o,\delta=35^o)$. That statistical significance is far higher than the 1.29$\sigma$ reported in Table 1 of \citep{iye2020spin}. Reasons for the differences are discussed in Section~\ref{reasons}.

A dipole axis formed by the distribution of galaxy spin directions means that one hemisphere has a higher number of galaxies that spin in one direction, while that asymmetry is inverse in the opposite hemisphere. Table~\ref{hemispheres} shows the distribution of galaxy spin directions when separating the sky into two hemispheres.

\begin{table*}
\scriptsize
\begin{tabular}{|l|c|c|c|c|c|}
\hline
Hemisphere & \# Z-wise          & \# S-wise         &  $\frac{\#Z}{\#S}$  & P                  & P                   \\
   (RA)         &                        &                         &                               & (one-tailed)   & (two-tailed)     \\
\hline
$70^o-250^o$               & 23,037 & 22,442   &   1.0265   &  0.0026   &  0.0052     \\
$>250^o \cup <70^o$   &  13,660 &  13,749  &   0.9935   &  0.29      &  0.58         \\
\hline
\end{tabular}
\caption{The number of galaxies in the \citep{iye2020spin} catalogue that spin in opposite directions when separating the sky into two hemispheres.}
\label{hemispheres}
\end{table*}

Statistically significant non-random distribution is observed in one of the hemispheres. In the opposite hemisphere the asymmetry is not statistically significant, but the asymmetry is inverse to the asymmetry in the hemisphere centred at (RA=160$^o$). When applying statistical Bonferronni correction for the two hemisphere, the statistical significance is still $\sim0.0104$. A Monte Carlo simulation showed that the probability to have such distribution by chance is $P\simeq0.007$. Code and instructions to reproduce the analysis are provided at \url{https://people.cs.ksu.edu/~lshamir/data/iye_et_al}. The fact that a simple separation into two hemisphere is statistically significant shows that a method that shows random distribution of this specific dataset is incomplete.

The algorithm used to annotate the galaxies works by first converting the galaxy image into its radial intensity plot transformation, and then detecting peaks in the transformation to identify the shift in the peaks, which show the shift in the arms, and therefore the curve \citep{shamir2011ganalyzer}. To perform a correct identification of the spin direction of the galaxy, a sufficient number of peaks detected in the radial intensity plot is required. Therefore, if a galaxy does not have a certain minimum number of peaks detected in its radial intensity plot, that galaxy is rejected from the analysis.

The analysis shown in Figure~\ref{dipole1} is the result of using the galaxies used in \citep{shamir2017photometric}. That experiment aimed at performing a photometric analysis rather than an attempt to identify a dipole axis in the spin directions of the galaxies as was done in \citep{iye2020spin}. The minimum number of peaks detected in the radial intensity plot to determine the spin direction of a galaxy in \citep{shamir2017photometric} was 10 peaks. That is, if after converting the galaxy image to its radial intensity plot 10 peaks or more were detected, the spin direction of the galaxy could be determined based on these peaks. The low number of peaks increased the number of annotated galaxies, but also led to a certain inaccuracy of the galaxy annotation. The inaccuracy of the annotations is discussed in \citep{shamir2017photometric}. In the previous experiments of identifying a dipole axis in the spin directions of the galaxies, the minimum number of peaks required to make an annotation was 30 \citep{shamir2020patterns}. The certain inaccuracy of the annotations of the galaxy images can lead to weaker signal compared to previous studies with a similar number of galaxies such as \citep{shamir2020patterns}. Another reason for the weaker signal is that the objects used in \citep{shamir2017photometric} were relatively bright objects $(i<18)$, and therefore objects with lower redshift. But despite these selection criteria, the signal observed in the experiment is stronger than 2$\sigma$.



\subsection{Reasons for the differences in the results}
\label{reasons}

\cite{iye2020spin} state that the ``clean'' dataset of 72,888 galaxies exhibits random distribution of 1.29$\sigma$ in the spin directions of the spiral galaxies. These results are in conflict with the results shown in Section~\ref{results}, showing that the reproduction of the analysis with the exact same data shows a much stronger statistical significance, stronger than 2$\sigma$. One reason that can lead to lower statistical significance is that \cite{iye2020spin} used the photometric redshift to limit the volume of the galaxies. For instance, the 0.29$\sigma$ highlighted in the abstract was determined after using the photometric redshift. The \citep{iye2020spin} paper uses the term ``measured redshift'', but as explained in \citep{shamir2017photometric} the vast majority of the galaxies in that dataset do not have spectra, and therefore the galaxies could not have spectroscopic redshift. As stated in the journal version of \citep{iye2020spin}, the source of the redshifts is the catalogue of \citep{paul2018catalog}, which is a catalogue of photometric redshift. The photometric redshift is highly inaccurate and can be systematically biased. The inaccuracy of the photometric redshift can add substantial bias to the results, and can lead to lower statistical signal due to the unexpected inaccuracies added to the data.

However, the analysis when using the entire ``clean'' dataset was done without limiting the volume, and without using the photometric redshift. Therefore, the photometric redshift cannot be the reason for the discrepancy between the reported and observed results. An inquiry to the National Astronomical Observatory of Japan (NAOJ), where the research was conducted, provided a reason for the differences. The analysis is available at \citep{Fukumoto2021}. The full analysis of the NAOJ can be accessed at \url{https://people.cs.ksu.edu/~lshamir/data/iye_et_al/watanabe_NAOJ_reply.pdf}.

According to the analysis of the NAOJ \citep{Fukumoto2021}, the analysis was done by assuming that the galaxies are distributed uniformly in the hemisphere.  As the analysis summary states: ``Because it is hard to verify the detail of simulations, we here calculate the analytic solution by Chandrasekhar (1943) which assumes uniform samples in the hemisphere.'' That is, the statistical significance can be reproduced to a certain extent if assuming that the galaxies used in the analysis are distributed uniformly in the hemisphere. When making that assumption, the statistical signal of the dipole axis is 1.35$\sigma$, which is close to the 1.29$\sigma$ reported in \citep{iye2020spin}. 

\begin{figure}
\centering
\includegraphics[scale=0.66]{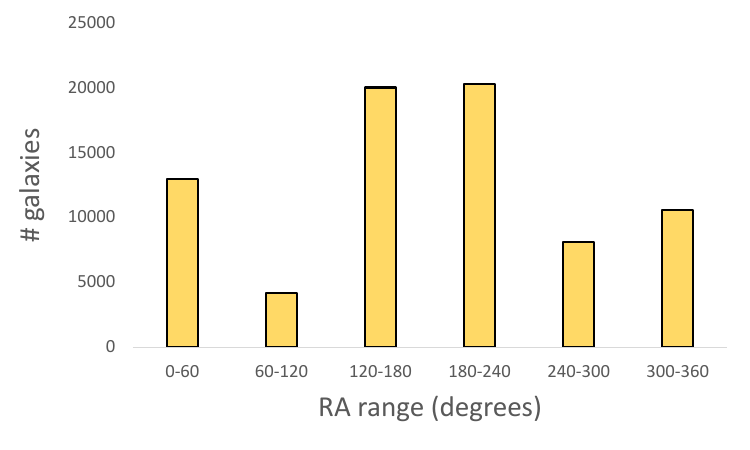}
\caption{RA distribution of the galaxies.}
\label{ra_distribution}
\end{figure}

\cite{iye2020spin} do not mention in the paper the analytic solution of Chandrasekhar (1943), or the assumption of a uniform sample in the hemisphere. 
More importantly, the assumption of uniform distribution in the hemisphere is not true for SDSS in general, and specifically not true for the dataset used here. for instance, Figure~\ref{ra_distribution} shows the distribution of the RA of the galaxies. As the figure shows, the distribution is not uniform or close to uniformity, and some RA ranges are far more populated than other ranges.

To transform into a uniform sample, the locations of the galaxies need to change so that they are spread uniformly. These changes in the locations can lead to changes in the results of the analysis of the dipole axis. For instance, the file \url{https://people.cs.ksu.edu/~lshamir/data/iye_et_al/galaxies_uniform.csv} is the same 72,888 galaxies such that their RA values are distributed uniformly in the range of $(0,360)$, and the declination values are distributed uniformly in the declination range of the galaxies in the original dataset, which is $(-24^o,5,79^o)$. 

\begin{figure}
\centering
\includegraphics[scale=0.16]{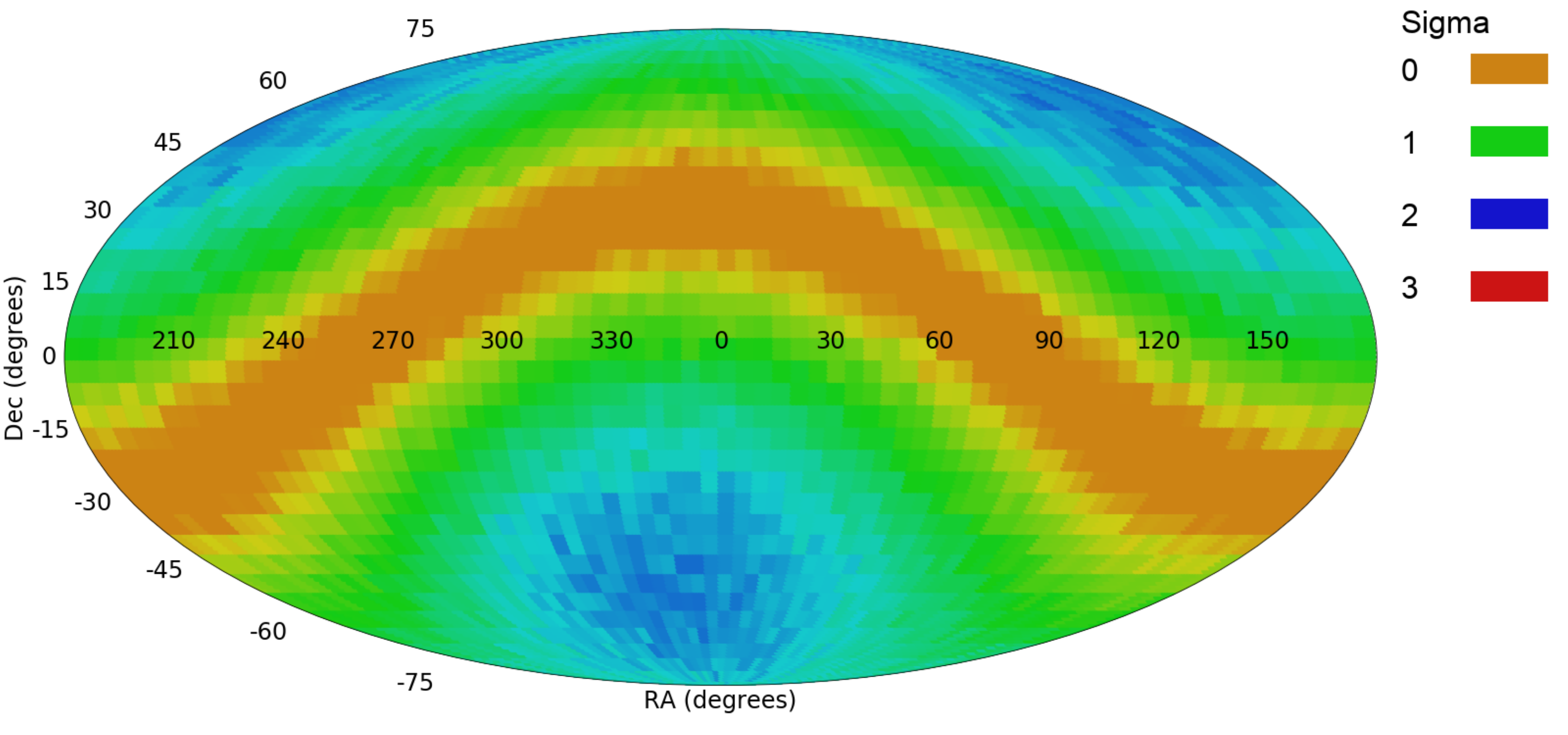}
\caption{The statistical significance of a dipole axis from all $(\alpha,\delta)$ combinations when the distribution of the RA and declination is uniform.}
\label{dipole_uniform_ra_dec}
\end{figure}

Figure~\ref{dipole_uniform_ra_dec} shows the same analysis as Figure~\ref{dipole1}, but with the uniformly distributed galaxies. The most likely dipole axis is identified at $(\alpha=172^o,\delta=51^o)$, with statistical significance of 1.68$\sigma$. That statistical signal is still higher than the 1.35$\sigma$ reported in \citep{Fukumoto2021}, but it is lower than the statistical signal observed when using the real locations of the galaxies, without making any assumption regarding the nature of their distribution in the hemisphere. But the weaker signal shows that the assumption that the galaxies are distributed uniformly leads to a weaker signal, and could be the reason for the weaker signal observed by \cite{iye2020spin}.

\section{The possibility of error in the galaxy annotation}
\label{error}

\begin{figure}[h]
\centering
\includegraphics[scale=0.15]{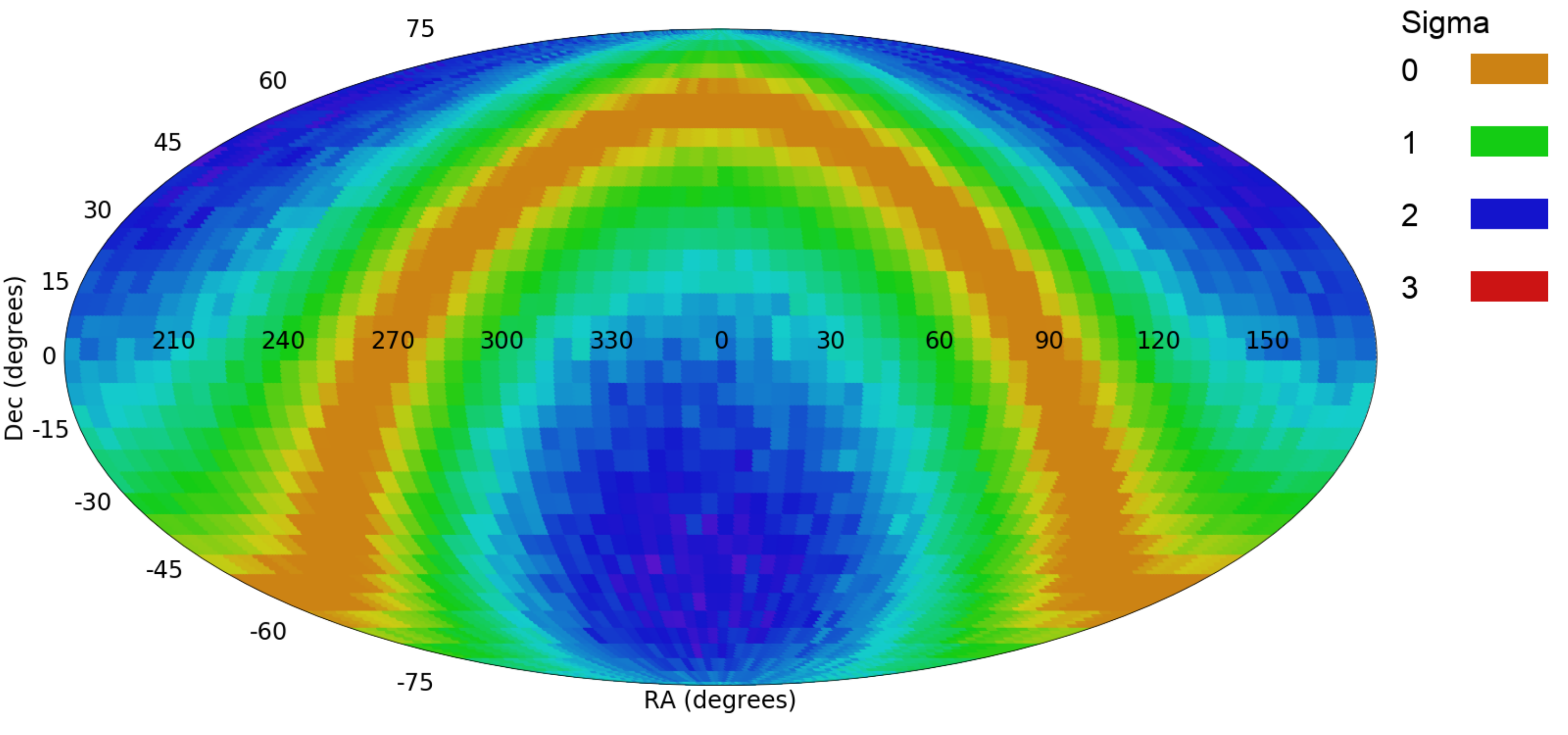}
\caption{The statistical significance a dipole axis to exist by chance from different $(\alpha,\delta)$ combinations after mirroring the images.}
\label{dipole1_mirrored}
\end{figure}

The {\it Ganalyzer} algorithm used to annotate the galaxies in Section~\ref{reproduction} is a simple model-driven algorithm that follows defined rules. It does not use machine learning or pattern recognition paradigms that their high complexity make them a ``black box'', and are very difficult to analyze and validate \citep{rudin2019stop,dhar2022systematic}. The simple ``mechanical'' nature of {\it Ganalyzer} allows it to ensure that the analysis is symmetric, as was shown experimentally in \citep{shamir2021large,shamir2021particles,shamir2022asymmetry,shamir2022analysis2}. The same analysis shown in Section~\ref{results} was repeated after mirroring all galaxies by using the {\it ImageMagick} ``flop'' command. Figure~\ref{dipole1_mirrored} shows the results, which is the same as Figure~\ref{dipole1}, and a very similar statistical signal of 2.17$\sigma$. The slight difference in statistical signal is expected due to the random assignment of spin directions.

Assuming that the annotation of the galaxies has a certain error, Equation~\ref{asymmetry} defines the asymmetry {\it A} between galaxies spinning in opposite  directions in a certain part of the sky as observed from Earth
\begin{equation}
A=\frac{(N_{cw}+E_{cw})-(N_{ccw}+E_{ccw})}{N_{cw}+E_{cw}+N_{ccw}+E_{ccw}},
\label{asymmetry}
\end{equation}
where $N_{cw}$ and $N_{ccw}$ are the numbers of galaxies spinning clockwise and counterclockwise, respectively, and $E_{cw}$ and $E_{ccw}$ are the numbers of galaxies incorrectly annotated as spinning clockwise and counterclockwise, respectively.

Because the number of galaxies incorrectly annotated as spinning clockwise is expected to be about the same as the number of galaxies incorrectly annotated as spinning counterclockwise. When $E_{cw} \simeq E_{ccw}$, {\it A} can be defined by Equation~\ref{asymmetry2}.

\begin{equation}
A=\frac{N_{cw}-N_{ccw}}{N_{cw}+E_{cw}+N_{ccw}+E_{ccw}}
\label{asymmetry2}
\end{equation}

Since $E_{cw}$ and $E_{ccw}$ must be positive integers, {\it A} gets lower as $E_{cw}$ and $E_{ccw}$ gets higher. Therefore, an error in the annotation algorithm can only make the asymmetry {\it A} smaller. The effect of incorrectly annotated galaxies was studied in \citep{shamir2021particles}. That analysis showed that when adding artificial error to the annotation in a symmetric manner the results do not change substantially, and the signal does not increase when the error is added. But when adding the error in a non-symmetric manner, even a small error of merely 2\% leads to a dipole the peaks in the celestial pole, with very high statistical significance \citep{shamir2021particles}.

\begin{figure}
\centering
\includegraphics[scale=0.74]{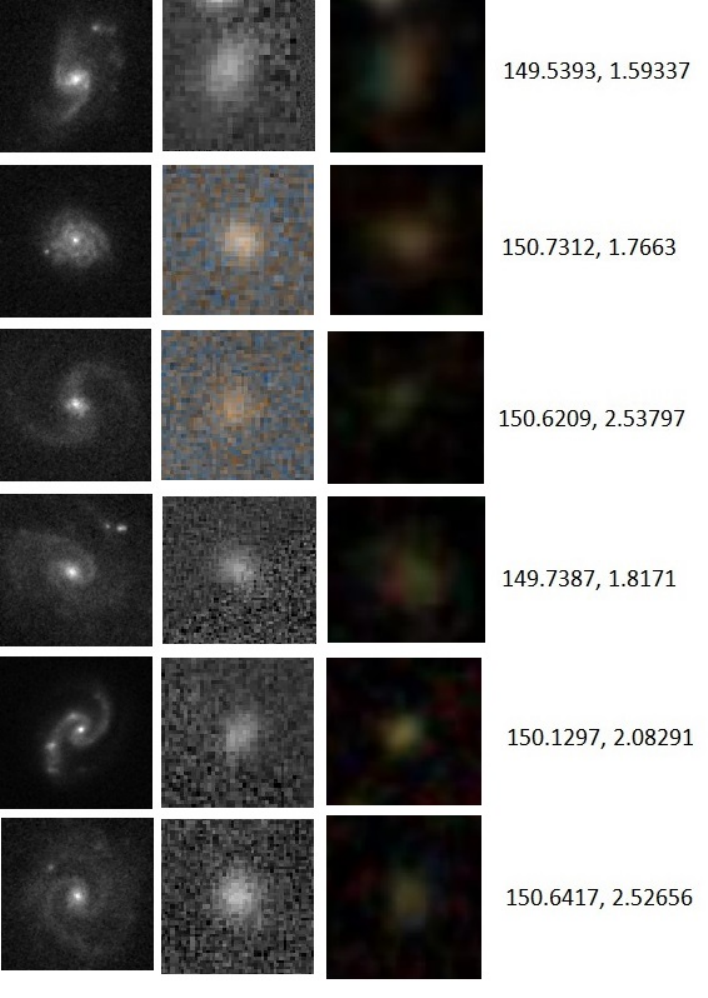}
\caption{Galaxies images by HST (left), Pan-STARRS (middle), and SDSS (right). The $(\alpha,\delta)$ coordinates of each galaxy are also specified. The images show that galaxies that have clear spin direction in HST cannot be annotated when using the Earth-based SDSS or Pan-STARRS.}
\label{sdss_hst}
\end{figure}

Another aspect is the completeness of the analysis. While some of the galaxies were annotated by their spin directions, most galaxies in the initial dataset were not assigned with a spin direction, and were therefore excluded from the analysis. These galaxies did not have an identifiable spin directions. Clearly, many of these galaxies do spin in a certain direction, but the direction cannot be determined from the image. For instance, Figure~\ref{sdss_hst} shows galaxies imaged by Pan-STARRS, SDSS, and HST.

As the figure shows, galaxies imaged by SDSS and Pan-STARRS do not have an identifiable spin direction, while the HST images of the same galaxies show that these galaxies have clear spin patterns. It is obvious that galaxies imaged by HST and do not have an identifiable spin patterns also cannot be assumed to have no spin direction, as HST also has a limiting magnitude. The symmetric nature of the algorithm is therefore critical to ensure the absence of a small bias that can lead to biased results. Other reasons that can affect the analysis are galaxies with leading arms, cosmic variance, hardware flaws, and atmospheric effect as discussed in Section 4 in \citep{shamir2022asymmetry} or in \citep{shamir2022analysis2}.

In all of these previously collected datasets, the direction of rotation of the galaxies were determined by the shape of the arms. Therefore, the galaxies are face-on galaxies, allowing an Earth-based observer to identify the arms and their curves.  Obviously, the datasets used here are all datasets used in previous experiments, and were used here in the same manner. These experiments did not include the inclination of the galaxies in the analysis, as these galaxies are mostly face-on galaxies, but it can be assumed that the inclination of the face-on galaxies in not exactly 90 degrees for all of them. But it is also expected that these inclination variations will be distributed equally between galaxies that spin clockwise an galaxies that spin counterclockwise. Therefore, if the expected slight inclination variations affect the analysis, it is expected to affect clockwise and counterclockwise galaxies in a similar manner, and therefore cannot lead to a preferred spin direction.

\section{Explanation to the observation that is not related to the large-scale structure}
\label{explanation}

One of the explanations to the observation described in this paper is that the observation reflects the real Universe. The contention that the Universe is oriented around a major axis shifts from the standard cosmological models. It might be in agreement with other theories such as ellipsoidal universe \citep{campanelli2006ellipsoidal,gruppuso2007complete}, rotating Universe \citep{godel1949example,ozsvath1962finite,ozsvath2001approaches}, and black hole cosmology \citep{pathria1972universe,stuckey1994observable,easson2001universe,seshavatharam2010physics,poplawski2010radial,christillin2014machian,dymnikova2019universes,chakrabarty2020toy,poplawski2021nonsingular,seshavatharam2022concepts,gaztanaga2022black,gaztanaga2022black2}, which assume the existence of a cosmological-scale axis, but it is not aligned with the standard model. On the other hand, it is also possible that the Universe does not have a cosmological-scale axis, and the observed asymmetry is driven by internal structure of galaxies \citep{shamir2020asymmetry,mcadam2023asymmetry}. In that case, the observation can be explained without the need to modify the standard cosmological model.

\begin{figure*}
\centering
\includegraphics[scale=0.3]{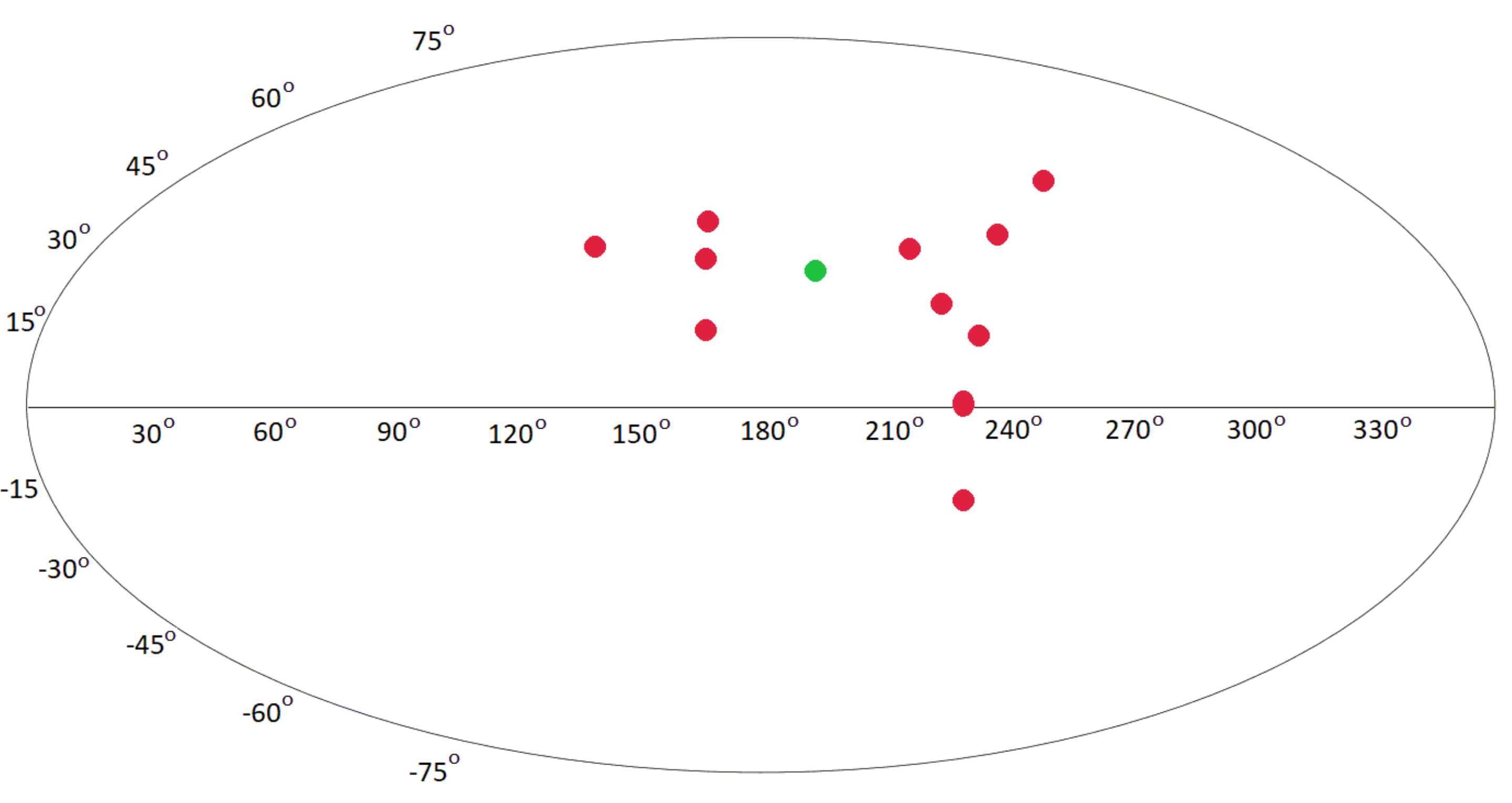}
\caption{The locations of the most likely dipole axes in several previous experiments \citep{land2008galaxy,longo2011detection,shamir2012handedness,shamir2016asymmetry,shamir2020patterns,shamir2021large,shamir2021particles,shamir2022asymmetry,shamir2022analysis2}, and the location of the galactic pole (green) at $(\alpha=192^o, \delta=27^o)$.}
\label{poles_position}
\end{figure*}

One of the indications that the observed asymmetry could be driven by internal structure of galaxies is that the peak of the dipole axes as determined by the different experiments and different telescopes tend to be close to the Galactic pole. Figure~\ref{poles_position} displays the peaks of the axes observed in 11 previous experiments using SDSS \citep{land2008galaxy,longo2011detection,shamir2012handedness,shamir2016asymmetry,shamir2020patterns,shamir2021particles}, Pan-STARRS \citep{shamir2020patterns}, DESI Legacy Survey \citep{shamir2022analysis2}, DES \citep{shamir2022asymmetry}, and DECam \citep{shamir2021large}. As the figure shows, the axes peak with proximity to the Galactic pole, leading to the possibility that the observation is driven by internal structure of galaxies and the rotation of the observed galaxies relative to the Milky Way. The analysis described in this paper also shows a dipole axis that peaks with close proximity to the galactic pole. That was observed with galaxies annotated by Ganalyzer, as well as the experiments with galaxies annotated by SPARCFIRE, as shown by Figure~\ref{sparcfire}.

Comparison of the brightness of these galaxies shows a statistically significant difference in the brightness of clockwise and counterclockwise galaxies. That was also shown with galaxies from SDSS, Pan-STARRS, HST, and DESI Legacy Survey \citep{shamir2020asymmetry,mcadam2023asymmetry}. The difference in galaxy brightness can be linked directly to the asymmetry in the number of galaxies that spin in opposite directions. Naturally, if galaxies around the galactic pole that rotate counterclockwise are slightly brighter than galaxies that rotate clockwise, more counterclockwise galaxies will be detected in that part of the sky. That will lead to asymmetry between the number of galaxies that spin in opposite directions, and a dipole axis that peaks around the galactic pole. That axis is not a feature of the large-scale structure, but driven by internal structure of galaxies.


To test that, it is possible to compare the brightness of galaxies that spin in opposite directions, and located around the Galactic pole. For instance, analyzing the exponential magnitudes of SDSS galaxies annotated automatically by Ganalyzer used in \citep{shamir2020patterns} provides 4,087 galaxies in the $50^o \times 50^o$ part of the sky around the Northern galactic pole. The dataset is available at \url{https://people.cs.ksu.edu/~lshamir/data/sdss_phot}. Table~\ref{sdss_dr7} shows the average exponential magnitude of galaxies spinning clockwise and galaxies spinning counterclockwise.

  

\begin{table}
\centering
\scriptsize
\begin{tabular}{lHHcccccc}
\hline
Band          & \# cw     & \# ccw    & Mag           & Mag  & $\Delta Mag$   & P t-test \\
                  &  galaxies & galaxies  &  cw             &  ccw  &                     & (two-tailed)  \\ 
\hline
G   & 2046  &  2041  &  17.3933$\pm$0.02       & 17.3217$\pm$0.02       &  0.0716 &  0.0112 \\
R &	2046  & 2041   &  16.8502 $\pm$0.02      &  16.7788$\pm$0.02       & 0.0714  & 0.0116 \\
Z   & 2046  &  2041  &  16.4213$\pm$0.02       & 16.34909$\pm$0.02        & 0.0722   &  0.01 \\
\hline
\end{tabular}
\caption{The magnitude of SDSS galaxies with different spin directions. All galaxies are in the $50^o \times 50^o$ window centred at the Northern galactic pole. The galaxies are annotated by {\it Ganalyzer}.}   
\label{sdss_dr7}
\end{table}

As the table shows, galaxies spinning counterclockwise in the part of the sky around the Northern galactic pole are brighter than galaxies spinning counterclockwise in the same part of the sky. These results are aligned with the results of similar experiments with other telescopes \citep{mcadam2023asymmetry}, and explain the higher number of counterclockwise galaxies observed in that part of the sky. In this case, the asymmetry in the number of galaxies can be attributed to galaxy rotation and internal structure of galaxies, rather than to the large-scale structure of the Universe. 

A similar analysis with the galaxies annotated by {\it Galaxy Zoo} also shows similar magnitude difference. Table~\ref{north_pole_galaxy_zoo} shows a similar analysis using all Galaxy Zoo galaxies in the $50^o \times 50^o$ window centred at the Northern galactic pole, that also met the ``superclean'' criterion of Galaxy Zoo. A galaxy annotation is considered ``superclean'' if 95\% or more of the votes agree on the annotation \citep{land2008galaxy}. That provided a dataset of 2,841 galaxies. The results show similar magnitude difference compared to the galaxies annotated by {\it Ganalyzer} as shown in Table~\ref{sdss_dr7}.



\begin{table}
\centering
\scriptsize
\begin{tabular}{lHHcccccc}
\hline
Band          & \# cw     & \# ccw    & Mag           & Mag  & $\Delta Mag$   & t-test P \\
                  &  galaxies & galaxies  &  cw             &  ccw  &                     &  (one-tailed)  \\ 
\hline
G &	1290  & 1551 &   16.3582$\pm$~0.03      &  16.2964$\pm$~0.03         & 0.0618  & 0.07 \\
R   & 1290  &  1551  &  15.7982$\pm$~0.03     & 15.7310$\pm$~0.03       &  0.0671 &  0.058 \\
Z   & 1290  &  1551  & 15.3972$\pm$~0.03      & 15.3184$\pm$~0.03        & 0.0788   &  0.032 \\
\hline
\end{tabular}
\caption{The magnitude of SDSS galaxies that spin in opposite directions as annotated by Galaxy Zoo in the $50^o \times50^o$ window centred at the Northern galactic pole. The galaxies include just galaxies that their annotation met the Galaxy Zoo ``superclean'' criterion.}   
\label{north_pole_galaxy_zoo}
\end{table}

The ``superclean'' criterion of Galaxy Zoo provides cleaner data, but also leads to the sacrifice of substantial number of galaxies that do not meet the criterion. Galaxy Zoo therefore has also the ``clean'' criterion, according which 80\% or more of the annotation needs to agree. Table~\ref{north_pole_galaxy_zoo_clean} shows the results with Galaxy Zoo ``clean'' galaxies, which provided a larger set of 9,512 galaxies. The results show that the absolute difference is smaller compared to using the ``superclean'' annotations, which can be explained by the higher number of incorrectly annotated galaxies in the ``clean'' annotations compared to the ``superclean'' annotations. The difference, however, is still statistically significant, also due to the higher number of galaxies that meet the ``clean'' criterion.

\begin{table}
\centering
\scriptsize
\begin{tabular}{lHHcccccc}
\hline
Band          & \# cw     & \# ccw    & Mag           & Mag  & $\Delta Mag$   & t-test P \\
                  &  galaxies & galaxies  &  cw             &  ccw  &                     &  (one-tailed)  \\ 
\hline
G & 4550	  & 4962 &   16.7475$\pm$~0.01      & 16.7183$\pm$~0.01         & 0.0292  & 0.019 \\
R   & 4550  &  4962  &  16.1833$\pm$~0.01     & 16.1573$\pm$~0.01       &  0.026 &  0.033 \\
Z   & 4550  &  4962  & 15.7566$\pm$~0.01      & 15.7287$\pm$~0.01        & 0.0279   &  0.024 \\
\hline
\end{tabular}
\caption{The magnitude of SDSS galaxies annotated by Galaxy Zoo in the $50^o \times50^o$ window centred at the galactic pole. The galaxies include just galaxies that their annotation met the Galaxy Zoo ``clean'' criterion.}   
\label{north_pole_galaxy_zoo_clean}
\end{table}


The brightness differences at the field around the Galactic pole shown in Tables~\ref{sdss_dr7} through~\ref{north_pole_galaxy_zoo_clean} can be compared to a control field perpendicular to the Galactic pole. Tables~\ref{sdss_dr7_control} and~\ref{north_pole_galaxy_zoo_clean_control} show the magnitude difference in the field centred at $(\alpha=102^o,\delta=0^o)$ with galaxies annotated by Ganalyzer and by Galaxy Zoo, respectively. The number of galaxies in these experiments are 3,781 and 5,094, respectively.

\begin{table}
\centering
\scriptsize
\begin{tabular}{lHHcccccc}
\hline
Band          & \# cw     & \# ccw    & Mag           & Mag  & $\Delta Mag$   & P t-test \\
                  &  galaxies & galaxies  &  cw             &  ccw  &                     & (one-tailed)  \\ 
\hline
G   & 1947  &  1834  &  17.4780$\pm$0.02       & 17.4678$\pm$0.02       &  0.01 &  0.36 \\
R &	1947  & 1834   &  16.8978$\pm$0.02      &  16.8920$\pm$0.02       & 0.006  & 0.42 \\
Z   & 1947  &  1834  &  16.4292$\pm$0.02       & 16.4154$\pm$0.02        & 0.014   &  0.31 \\
\hline
\end{tabular}
\caption{The magnitude of SDSS galaxies with different spin directions at a field perpendicular to the Galactic pole. The galaxies are annotated by {\it Ganalyzer}. The magnitude differences are far smaller compared to the magnitude difference in the field centred at the Galactic pole.}   
\label{sdss_dr7_control}
\end{table}



\begin{table}
\centering
\scriptsize
\begin{tabular}{lHHcccccc}
\hline
Band          & \# cw     & \# ccw    & Mag           & Mag  & $\Delta Mag$   & P t-test \\
                  &  galaxies & galaxies  &  cw             &  ccw  &                     & (one-tailed)  \\ 
\hline
G   & 2513  &  2581  &  16.849$\pm$0.02       & 16.8579$\pm$0.02       &  -0.009 &  0.37 \\
R &	2513  & 2581   &  16.2734$\pm$0.02      &  16.2768$\pm$0.02       & -0.003  & 0.45 \\
Z   & 2513  &  2581  &  15.82378$\pm$0.02       & 15.8224$\pm$0.02        & 0.0013   &  0.58 \\
\hline
\end{tabular}
\caption{The magnitude of SDSS galaxies with different spin directions at a field perpendicular to the Galactic pole. The galaxies are annotated by {\it Galaxy Zoo}.}   
\label{north_pole_galaxy_zoo_clean_control}
\end{table}



As both tables show, in the field perpendicular to the Galactic pole the brightness differences are far smaller, and statistically insignificant. Although the reason for the difference observed at around the Galactic pole is still unclear, it is possible that the motion of the observed galaxies relative to the Milky Way might be related to the observation. More work will be required to fully understand the reason for the observation.



\section{Conclusion}
\label{conclusion}

Several different probes have shown evidence of large-scale isotropy and parity violation \citep{aluri2022observable}. This paper aims at studying a probe with several studies that show conflicting conclusions by analyzing the studies and replicating the results. Code and data used for the analysis are available. The analyses and reproduction of the experiments show that the studies might not necessarily conflict with each other. 

While reproducibility is a key concept in science \citep{aristotle350}, it has been shown that more researchers tried but failed to reproduce work of their colleagues \citep{baker20161}. These failed attempts are normally left unknown to the public, and attempts to make them public through the scientific literature are often repelled through the peer-review process, leading to unbalanced information in the scientific literature \citep{baker20161}. In the case of computational analyses, reproducibility is expected to be more straightforward compared to ``wet'' laboratory experiments that may require highly trained researchers just to follow the protocols and reproduce the results. Yet, comprehensive analysis  have shown that even in the case of papers published in the most regarded outlets with strict reproducibility policies, 76\% of the published papers could not be reproduced \citep{stodden2018empirical}.

The attempt to reproduce the results and analyze the research aims at identifying the reasons that explain why different studies show conflicting conclusions. The results join a large number of other observations of cosmological-scale anisotropy reflected through multiple probes \citep{aluri2022observable}, although another explanation to the observation that does not require violation of the cosmological principle is also proposed. Other explanations are also possible.

Future studies with more data will be required to profile the nature of the observation. The Vera C. Rubin Observatory will provide the world's largest astronomical database, and will allow to provide analysis with high resolution that will help to fully profile the observation and understand its nature. It is expected that such analysis will identify the exact location of the peak of the axis, which might allow to associate it with other observations that can include the Galactic pole, the CMB Cold Spot, the CMB dipole, etc. Instruments such as the Dark Energy Spectroscopic Instrument \citep{martini2018overview} will provide spectra of a large number of galaxies. That will allow to study whether the asymmetry changes with the redshift. Early observations using $\sim 6.4 \cdot 10^4$ galaxies with spectra showed evidence that the asymmetry increases as the redshift gets higher \citep{shamir2020patterns}, suggesting that the asymmetry is of primordial origin. The Dark Energy Spectroscopic Instrument will allow to profile that change with an unprecedented number of galaxies with spectra.

\section*{Acknowledgments}

I would like to thank the three knowledgeable anonymous reviewers for the comments that helped to improve the paper. This study was supported in part by NSF grants AST-1903823 and IIS-1546079. 


\section*{Data Availability}

The data used in this paper is publicly available. URLs were the data get be accessed are available inside the manuscript. SDSS galaxies analyzed by {\it SPARCFIRE} are available at \url{https://people.cs.ksu.edu/~lshamir/data/sparcfire/}. The data used in \citep{shamir2022using} are available at \url{https://people.cs.ksu.edu/~lshamir/data/iye_et_al}. Data for comparing the brightness of SDSS galaxies around the galactic pole is available at \url{https://people.cs.ksu.edu/~lshamir/data/sdss_phot}.

\bibliographystyle{apalike}
\bibliography{main_arxiv}

\end{document}